\documentstyle[preprint,tighten,aps]{revtex}

\def\Rvet{\bf R}
\def\enne{\bf n}

\begin{document}

\preprint{\vbox{\null\hfill INFNFE-01-97}}

\title{How long does it take for heat to flow through the sun?}

\author{G.~Fiorentini$^{\S,\dag}$
         and B.~Ricci$^{\dag}$
 }

\address{
$^{\S}$Dipartimento di Fisica dell'Universit\`a di Ferrara,
       via Paradiso 12, I-44100 Ferrara, Italy\\
$^{\dag}$Istituto Nazionale di Fisica Nucleare, Sezione di Ferrara,
      via Paradiso 12, I-44100 Ferrara, Italy\\
}

\date{February 1997}

\maketitle

~\\

{\bf This question, which often comes in when discussing  solar neutrinos,
is not clearly analysed in any textbook, to our knowledge.
In this note we give a simple estimate 
of the flow time, also reminding its (ir)relevance
to the solar neutrino problem, see\cite{libro}.}

Most of the time is spent to flow from the energy
production region, say $R\leq R_e= 0.2\,R_\odot$, to the bottom
of the convective envelope, located at $R_b=0.7\,R_\odot $, as 
radiative transfer is much less efficient than
the  convective one.

As a simplification, we assume that in the radiative region
each photon makes a  random walk with mean free path $\lambda=1/ \kappa \rho$,
where $\rho(R)$ and $\kappa(R)$ are the solar density and opacity
calculated according to the standard solar model, see {\em e.g.} \cite{Ciacio}.
The radial dependence of $\lambda$ has to be
taken into account, since $\lambda$ increases by a factor twenty
when moving from $R_e$ to $R_b$, see Fig. \ref{fig1}.

After a time  $\Delta t= \lambda/c$, the position $\Rvet$ of the
photon is changed to:
\begin{equation}
\label{erre}
\Rvet {\mbox{$ (t+\Delta t)$}}= 
\Rvet {\mbox{$(t)$}} + \lambda {\mbox{$(R)$}} \enne \quad ,
\end{equation}
where $\enne$ is a unit random vector.
By squaring both sides of Eq. (\ref{erre}) and differentiating with respect
to time one has:
\begin{equation}
\label{evol1}
\frac{d}{dt} R^2 =  c\lambda (R)  + 2c\Rvet \cdot \enne \quad .
\end{equation}
The last term vanishes  when averaging  over the
photons isotropic ensemble:
\begin{equation}
\label{evol2}
\frac{d}{dt} \langle R^2 \rangle= c \langle\lambda (R) \rangle  
\end{equation}
By the additional approximation $\langle \lambda (R) \rangle=
\lambda(\langle R^2 \rangle ^{1/2}) $, Eq. (\ref{evol2}) can be
immediately integrated:
\begin{equation}
\label{integral}
t(R_1,R_2)= \frac{1}{c} \int_{R_1}^{R_2} \frac{2xdx}{\lambda(x)} \quad .
\end{equation}

From Fig. \ref{fig1} one sees that most of the time is spent
in the inner regions, where opacity is higher. From the same
figure one derives  $t(R_e,R_b)=4\cdot10^4$ yr.

As well known, for a constant mean free path $\lambda_0$ one has
$t(R_e,R_b)= (R_b^2- R_e^2)/(c \lambda_0)$.
By taking $\lambda_0=50\,\mu$m, as in the solar center, one would
overestimate the flow time  by an order of magnitude. 
On the other hand, for a typically quoted mean free path
$\lambda_0 \simeq 1$ cm, the time is
underestimated by an order of magnitude.
All this shows the relevance of the radial dependence 
of $\lambda$.

The deficit of solar neutrinos
 might be related to
a hypothetical solar instability,
such that the present nuclear energy production rate $\epsilon _\odot$ 
is smaller than the observed luminosity $L_\odot$. 
One is tempted to identify $t(R_e,R_b)$ as the time scale
over which the  difference between $\epsilon_\odot$ and $L_\odot$
is undetectable.
Actually the pertinent time scale is a much longer one. 

After the fire is switched off, the cooling time
of a pot is not determined by the phonon propagation time,
but by its thermal capacity. 
Much in the same way, if nuclear reactions were switched off
the sun would keep its present luminosity as long as it can be substained
by the thermal and gravitational energy $U_\odot$ stored in it, 
i.e. for a time $t_{HK}=U_\odot/L_\odot \simeq 3\cdot 10^7$ yr, as predicted 
by Helmholtz and Kelvin long ago, see also \cite{libro}.

\begin{figure}
\caption[q] {The   mean free path $\lambda$ (dashed line, right scale)
and the diffusion time T (full line, left scale) for  photons produced
at the solar center to reach the distance R, {\em i.e.} T(R)=t(0,R). }
\label{fig1}
\end{figure}

\end{document}